\documentclass{aa}  

\usepackage{graphicx}
\usepackage{booktabs}
\usepackage{txfonts}
\usepackage{color}
\usepackage{hyperref}
\usepackage{tabularx}
\usepackage{tikz}
\usepackage{tikz-feynman}
\usepackage{caption}
\usepackage{subcaption}
\usepackage{xcolor}
\usepackage{soul}
\usepackage[left]{lineno} 
\usepackage{newtxtext,newtxmath}
\usepackage{amsmath}	

\DeclareRobustCommand{\VAN}[3]{#2}
\let\VANthebibliography\thebibliography
\def\thebibliography{\DeclareRobustCommand{\VAN}[3]{##3}\VANthebibliography}

\newcommand{\nuc}[2]{$\mathrm{^{#2}#1}$}

\usepackage[normalem]{ulem}
\begin{document}

   \title{Search for Axion-Like Particles from Nearby Pre-Supernova Stars}


   \author{Saurabh Mittal
          \inst{1}
          \and
          Thomas Siegert
          \inst{1}
          \and
          Francesca Calore
          \inst{2}
          \and
          Pierluca Carenza
          \inst{3}
          \and
          Laura Eisenberger
          \inst{1}
          \and
          Maurizio Giannotti
          \inst{4,5}
          \and
          Alessandro Lella
          \inst{6,7}
          \and
          Alessandro Mirizzi
          \inst{6,7}
          \and
          Dimitris Tsatsis
          \inst{1}
          \and
          Hiroki Yoneda
          \inst{1,8,9,10,11}
          }

   \institute{Julius-Maximilians-Universität Würzburg, Fakultät für Physik und Astronomie, Institut für Theoretische Physik und Astrophysik,\\
   Lehrstuhl für Astronomie, Emil-Fischer-Str. 31, D-97074 Würzburg, Germany\\
   \email{saurabh.mittal@uni-wuerzburg.de}
   \and
   LAPTh, CNRS,  USMB, F-74940 Annecy, France
   \and
   The Oskar Klein Centre, Department of Physics, Stockholm University, Stockholm 106 91, Sweden
   \and
   Centro de Astropart{\'i}culas y F{\'i}sica de Altas Energ{\'i}as (CAPA), Universidad de Zaragoza, Zaragoza, 50009, Spain
   \and
   Physical Sciences, Barry University, 11300 NE 2nd Ave., Miami Shores, Florida 33161, USA
   %
   \and
   Istituto Nazionale di Fisica Nucleare—Sezione di Bari, Via Orabona 4, 70126 Bari, Italy
   \and
   Dipartimento Interuniversitario di Fisica Michelangelo Merlin, Via Amendola 173, 70126 Bari, Italy
   \and
   The Hakubi Center for Advanced Research, Kyoto University, Yoshida Ushinomiyacho, Sakyo-ku, Kyoto
    606-8501, Japan
   \and
   Department of Physics, Kyoto University, Kitashirakawa Oiwake-cho, Sakyo-ku, Kyoto 606-8502, Japan
   \and
   RIKEN Nishina Center, 2-1 Hirosawa, Wako, Saitama 351-0198, Japan
   \and
   Kavli Institute for the Physics and Mathematics of the Universe (WPI), UTIAS, The University of Tokyo,
5-1-5 Kashiwanoha, Kashiwa, Chiba 277-8583, Japan
   }

   \date{Received XXX; accepted YYY}

 
  \abstract
  {Axion-like particles (ALPs) are hypothetical pseudoscalar bosons that arise in many extensions of the Standard Model and can be well-motivated dark matter candidates. Nearby massive stars in the late stages of stellar evolution provide a promising environment for enhanced ALP production due to their high core temperatures and densities.}
  {In this work, we search for a combined signal of ALP-induced hard X-ray and soft $\gamma$-ray emission from 18 nearby pre-supernova stars. We use the full public INTEGRAL/SPI 22-year database to create individual datasets and link the resulting spectra for a coherent analysis. }
  {We use a maximum likelihood approach to extract the fluxes of the selected nearby stars from 20--2000\,keV. From stellar evolution models, we obtain the expected spectral shapes of ALPs producing processes 
  peaking in the range between 50--500\,keV, depending on the age and mass of the star. We then construct a joint likelihood that acknowledges the uncertainties in individual stellar parameters towards a combined estimate for the coupling constants $g_{a\gamma}$ and $g_{ae}$ as a function of the ALP mass $m_a$.}
  {We find that the hard X-ray and soft $\gamma$-ray fluxes of all selected stars are consistent with zero within uncertainties. We provide upper limits on the continuum flux as well as the 511\,keV and 1809\,keV lines from these sources. The combined estimate on the upper limit of the product $g_{a\gamma} \times g_{ae}$ is $(0.008 - 2) \times 10^{-24}~\mathrm{GeV}^{-1}$ (95\% C.I.) and the ALP-photon coupling $g_{a\gamma} = (0.13 - 1.26) \times 10^{-11}~\mathrm{GeV}^{-1}$ (95\% C.I.) up to a mass of $m_a \leqq 10^{-11}$ eV for different times to core-collapse and different magnetic field models.}
  {Our results are among the strongest limits on the ALP coupling constants in the literature. We also provide conservative limits on the coupling constants $g_{a\gamma} \times g_{ae}$ of $(0.27 - 1.25) \times 10^{-24}~\mathrm{GeV}^{-1}$ (95\% C.I.) by assuming all stars but one to be in the early He burning phase. This work shows that soft $\gamma$-ray observations are required to efficiently probe the ALP parameter space, as well as massive star evolution models in general.}


   \maketitle
%

\section{Introduction}\label{sec:intro}
Axion-like particles (ALPs), including the special case of QCD axions, are hypothetical light pseudo-scalar particles that couple very weakly to photons and other standard model particles. 
Their interactions are described by the Lagrangian: 
\begin{equation}
    \mathcal{L}_{\mathrm{int}} = -\frac{1}{4}g_{a\gamma}aF_{\mu\nu}\tilde{F}^{\mu\nu} - \sum_{f = e,p,n} g_{af} a \bar{f}\gamma_5 f,
    \label{eqn:lagrangian}
\end{equation}
where $F_{\mu \nu}$ is the electromagnetic field strength tensor, $\tilde{F}^{\mu \nu}$ its dual, $a$ the ALP field, and $f$ the SM fermion fields (in our cases, we will be interested only in electrons, $e$). 
The coupling constants $g_{a\gamma}$ and $g_{af}$ quantify the interaction strengths. 

QCD axions were originally introduced to resolve the strong CP problem in quantum chromodynamics and are characterized by a model-dependent relationship between their mass and coupling constants \citep{di2020landscape}.
More general ALPs emerge in several extensions of the Standard Model (SM) of particle physics \citep{Jaeckel:2010ni,ringwald2014axionsaxionlikeparticles,Agrawal:2021dbo,Giannotti:2022euq,Antel:2023hkf} and lack any specific relation among couplings and mass. 

From a top-down perspective, string theory predicts the presence of an ``axiverse'' with the QCD axion \citep{Peccei:1977hh,Peccei:1977ur,Weinberg:1977ma,Wilczek:1977pj} and several ultralight ALPs \citep{Arvanitaki:2009fg,Cicoli:2012sz,Cicoli:2023opf}.
From a bottom-up perspective, ALPs offer an interesting physics case in relation to dark matter \citep{Abbott:1982af,Dine:1982ah,Preskill:1982cy,Arias:2012az,Adams:2022pbo} and to several astrophysical puzzles \citep{Giannotti:2015kwo,giannotti2017stellar,Galanti:2022chk}.
In this context, stars have long been recognized as ALP factories \citep{Raffelt:1996wa,Raffelt:1999tx,Carenza:2024ehj},
as stellar plasmas provide ideal conditions for producing large fluxes of these particles.
The production of ALPs under these extreme temperature and density conditions, provides an additional channel for energy loss. 
This can alter the evolution of horizontal branch stars \citep{ayala2014revisiting,straniero2015axion,dolan2022advancing}, red giants \citep{capozzi2020axion, straniero2020rgb}, and white dwarfs \citep{bertolami2014WD}. 
Moreover, ALPs have been proposed as a possible explanation for the so-called stellar cooling anomaly—the observed excess cooling in several classes of stars \citep{giannotti2017stellar,Giannotti:2015kwo}. 
Additionally, they can also be used as supernova (SN) probes  since they are able to exit the stellar interiors earlier than photons \citep{Lella:2024hfk}. 
A comprehensive overview of the astrophysical implications of ALPs can be found in~\citet{Carenza:2024ehj}.

Notably, just a few years after the
introduction of the QCD axion, Pierre Sikivie \citep{Sikivie:1983ip} proposed to search for this elusive particle through dedicated observations of the Sun via the ``helioscopes'' technique~\citep{vanBibber:1988ge}.
The key idea of these types of experiments is that in the case of a ALP-photon coupling $g_{a\gamma}$, ALPs can be produced in the Sun's core via Primakoff process \citep{carlson1995pseudoscalar,primakoff1951photo}, and then convert into X-rays  in the magnetic field of the detector~\citep{CAST:2007jps}.
The CAST experiment, the most mature example of axion helioscope~\citep{Cetin:2024lzx}, has recently presented a new analysis \citep{CAST:2024eil}, improving the previous bound from Solar ALPs down to
$g_{a\gamma}< 5.7 \times 10^{-11} \,\ \textrm{GeV}^{-1}$ for $m_a \lesssim 0.02$~eV.
The next generation of helioscopes, BabyIAXO~\cite{IAXO:2020wwp,IAXO:2024wss} and the full scale IAXO~\cite{IAXO:2019mpb,IAXO:2025ltd}, aim to improve the sensitivity to the axion–photon coupling relative to CAST by a factor of 3 and by more than an order of magnitude, respectively.

The Sun is the closest star to us, so one would expect it to be the best target for astrophysical ALPs searches.
However, other stellar environments have been shown to have a competitive physics potential.
Notably, after the SN1987A neutrino observations, the occurrence of ALP burst produced in the SN core, simultaneously with neutrinos, was searched for. 
Notably, SN ALPs would have led to a gamma-ray burst, induced as a consequence of ALP-photon conversions in the Galactic magnetic field.
The non-observation of {such a} signal in the Gamma-Ray Spectrometer (GRS) of the SMM (Solar Maximum Mission) in coincidence the neutrino signal from SN 1987A,
provided a strong bound on ALPs coupled to photons \citep{Grifols:1996id,Brockway:1996yr, Hoof:2022xbe}.
For $m_a < 4 \times 10^{-10}$~eV it was found   $g_{a\gamma} < 5.3 \times 10^{-12}$ GeV$^{-1}$ \citep{Payez:2014xsa}.

Other promising stellar sources for ALPs are nearby red supergiant (RSG) stars.
Indeed, their high core temperatures and the steep dependence of the ALP production rate on temperature make them compelling sources of stellar ALPs.
Remarkably, there are $\sim 20$ supergiants with masses ranging from $10-30~ M_{\odot}$ within a distance $d\lesssim 1$ kpc.

\begin{figure}[htbp]
    \centering

    \begin{subfigure}[b]{0.5\textwidth}
        \centering
        \begin{tikzpicture}
            \begin{feynman}
                \vertex (a) {\(\gamma\)};
                \vertex[right=1.5cm of a] (b);
                \vertex[below=1.5cm of a] (c) {\(e,Ze\)};
                \vertex[right=1.5cm of c] (d);
                \vertex[right=1.5cm of d] (e) {\(e,Ze\)};
                \vertex[right=1.5cm of b] (f) {\(a\)};
                \node at (b) [anchor=north west, inner sep=2pt] {\(g_{a\gamma}\)};

                \diagram* {
                    (a) -- [photon] (b) -- [boson] (d),
                    (c) -- [fermion] (d),
                    (d) -- [fermion] (e),
                    (b) -- [scalar] (f),
                };
            \end{feynman}
        \end{tikzpicture}
        \caption{Primakoff}
    \end{subfigure}

    \par\medskip
    
    \begin{subfigure}[b]{0.45\linewidth}
        \centering
        \begin{tikzpicture}
            \begin{feynman}
                \vertex (i1) {\(e\)};
                \vertex[right=1.5cm of i1] (a);
                \vertex[below=1.5cm of i1] (i2) {\(\gamma\)};
                \vertex[right=1.5cm of i2] (b);
                \vertex[right=1.5cm of a] (f1) {\(a\)};
                \vertex[right=1.5cm of b] (f2) {\(e\)};
                \node at (a) [anchor=north west, inner sep=2pt] {\(g_{ae}\)};

                \diagram* {
                    (i1) -- [fermion] (a) -- [scalar] (f1),
                    (i2) -- [photon] (b) -- [fermion] (f2),
                    (a) -- [fermion] (b),
                };
            \end{feynman}
        \end{tikzpicture}
        \caption{Compton}
    \end{subfigure}
    \hfill
    \begin{subfigure}[b]{0.45\linewidth}
        \centering
        \begin{tikzpicture}
            \begin{feynman}
                \vertex (i1) {\(e\)};
                \vertex[below=1.5cm of i1] (i2) {\(Ze\)};
                \vertex[right=1.5cm of i1] (a);
                \vertex[right=0.3cm of a] (c);
                \vertex[right=1.5cm of i2] (b);
                \vertex[right=1.5cm of a] (f1) {\(e\)};
                \vertex[right=1.5cm of b] (f2) {\(Ze\)};
                \vertex[above=1cm of f1] (f3) {\(a\)};
                \node at (c) [anchor=north west, inner sep=2pt] {\(g_{ae}\)};

                \diagram* {
                    (i1) -- [fermion] (a) -- [fermion] (f1),
                    (i2) -- [fermion] (b) -- [fermion] (f2),
                    (a) -- [boson] (b),
                    (c) -- [scalar] (f3),
                };
            \end{feynman}
        \end{tikzpicture}
        \caption{bremsstrahlung}
    \end{subfigure}

    \caption{Feynman diagrams for ALPs production: Primakoff, Compton, and bremsstrahlung.}
    \label{fig:feynman}
\end{figure}
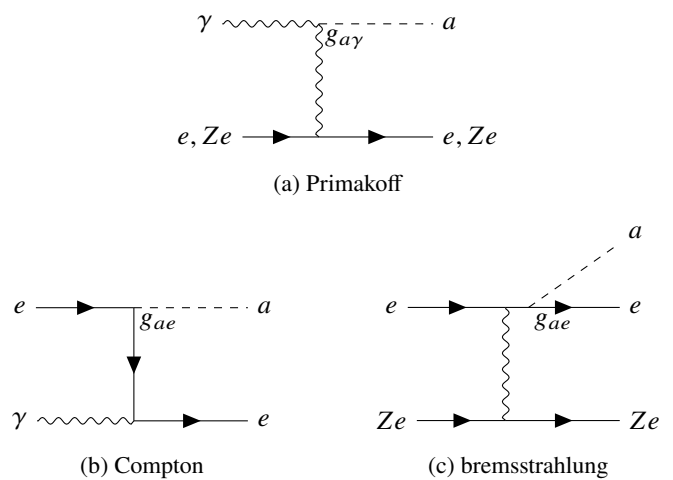

An example studied quantitatively is
the red supergiant star Betelgeuse ($\alpha$ Orionis), spectral type M2Iab at a distance $d \simeq 197$ pc \citep{dolan2016evolutionary}, proposed as ALP target in a seminal paper by E.~Carlson \cite{carlson1995pseudoscalar}.
Recently, \citet{Xiao:2020pra} used the data of a dedicated 50~ks observation by the NuSTAR satellite \citep{NuSTAR:2013yza} to place a 95\% C.I. upper limit on the ALP-photon coupling $g_{a\gamma} < (0.5 - 1.8) \times 10^{-11}$ GeV$^{-1}$ for ALP masses $m_{a} < (5.5 - 3.5) \times 10^{-11}$ eV, assuming only Primakoff production.
Enlarging the production channels to include, besides the Primakoff process, also  bremsstrahlung and Compton processes induced by the ALP-electron coupling $g_{ae}$, \citet{xiao2022betelgeuse} derived the constraint $g_{a\gamma} \times g_{ae}< (0.4-2.8)\times10^{-24}$ \textrm{GeV}$^{-1}$ for masses ${m_{a}\leq(3.5-5.5)\times10^{-11}}$ eV.
    Fig.~\ref{fig:feynman} shows the Feynman diagrams for the three ALP production mechanisms.

        The search we perform in this work is complementary to the study by \citet{xiao2022betelgeuse}.
        %
        Here, we extend their efforts to the $\gamma$-ray regime ($20 - 2000~\mathrm{keV}$) for Betelgeuse and 17 additional nearby ($< 1~\mathrm{kpc}$) red super giants using data from INTEGRAL/SPI \citep{winkler2003integral,vedrenne2003spi}.
        This paper is structured as follows: In Sec.\,\ref{sec:source_selection}, we describe our selection of candidate stars. 
        Sec.\,\ref{sec:spectrum} recapitulates on the expected spectral signatures from ALPs.
        Our data analysis method is described in Sec.\,\ref{sec:dataset_analysis}. 
        We present our results in Sec.\,\ref{sec:results} and conclude with an outlook in Sec.\,\ref{sec:conclusion}.
        
        \section{Source selection}\label{sec:source_selection}
	\begin{figure}[ht]
		\includegraphics[width=\linewidth]{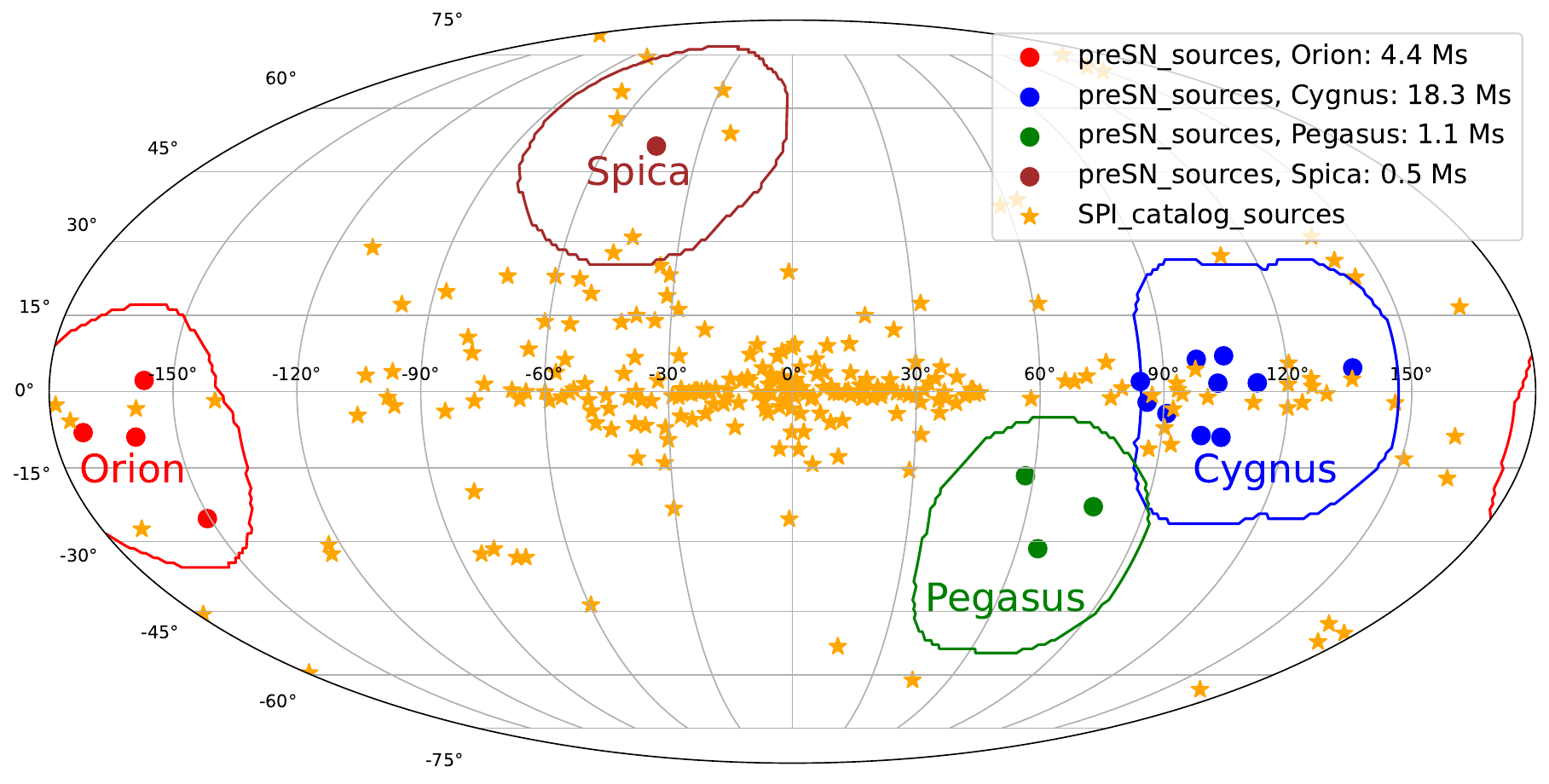}
		\centering
		\caption{The orange stars show the location of all the detected sources with SPI so far \citep{bouchet2008integral}. The circular dots are the red super giants used in this work. The colored boundaries are the exposure regions for each dataset.}
		\label{fig:Multiple_sources}
	\end{figure}
	The star sample chosen is a subset of the catalog provided in \citet{mukhopadhyay2020presupernova}.
    The catalog originally lists 31 core-collapse SN progenitors within 1 kpc that have both distance and mass estimates. 
    These massive stars are believed to be in the late stages of nuclear burning close to core-collapse.
    We shortlist 18 red super giant candidates from this list.
    In Tab.~\ref{tab:preSN_sources}, we present the details of the selected sources including their Galactic coordinates, common names, mass and distance.
	The stars are selected in a way that they are at large enough distances from bright sources in the SPI catalog (no bright sources within a $30^\circ$ radius of our candidate source) and from potentially variable sources to avoid systematics from source confusion and variability.

	Our final star sample consists of Betelgeuse (in close proximity with the Crab nebula, the brightest $\gamma$-ray source seen by INTEGRAL, and partly in SPI's field-of-view), CE Tauri (also close to the Crab nebula and Betelgeuse), Rigel (Orion region), ten stars in the Cygnus region (carefully selected to avoid Cyg X-1), three stars in the Pegasus region, Spica, and S Monocerotis A (Orion region) (Fig.~\ref{fig:Multiple_sources}).\\
	
\section{ALP spectrum and stellar models}\label{sec:spectrum}
\begin{figure}[ht!]
		\includegraphics[width=\linewidth]{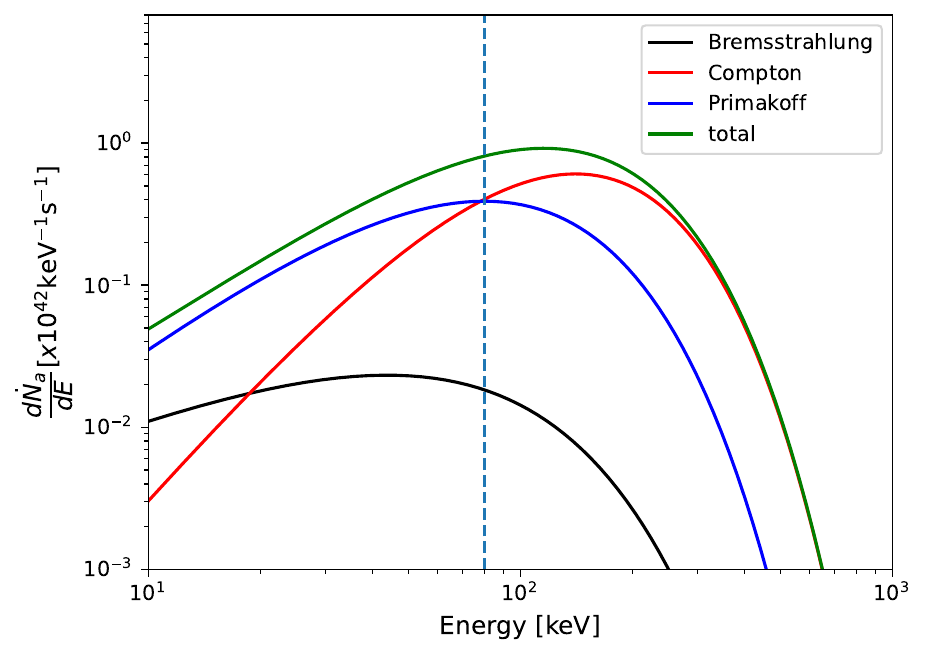}
		\centering
		\caption{Expected ALP fluxes from bremsstrahlung (black), Compton (red), Primakoff (blue), and total (green) using $g_{ae} = 10^{-13}$, $g_{a\gamma} = 10^{-11}~\mathrm{GeV}^{-1}$, and a time to core collapse of $\mathrm{t_{cc}} = 6900~\mathrm{yr}$. The blue dashed line shows the energy upper limit of NuSTAR ($79$~keV).}
		\label{fig:ALP_spectrum}
	\end{figure}

	The ALP source spectrum from a red super giant star can be approximated by
	\begin{equation}\label{eq:ALP_spectrum}
		\begin{aligned}
			\dfrac{d\dot{N}_a}{dE} = \dfrac{10^{42}}{\mathrm{keV~s}}
			\bigg[&C^Bg^2_{13}\left(\dfrac{E}{E_0^B}\right)^{\beta^B}e^{-(\beta^B + 1)E/E_0^B} \\
		 	+ &C^Cg^2_{13}\left(\dfrac{E}{E_0^C}\right)^{\beta^C}e^{-(\beta^C + 1)E/E_0^C} \\
				+ &C^Pg^2_{11}\left(\dfrac{E}{E_0^P}\right)^{\beta^P}e^{-(\beta^P + 1)E/E_0^P}\bigg]\mathrm{,}
		\end{aligned}
	\end{equation}
	where $g_{11} = g_{a\gamma} / 10^{-11}~\mathrm{GeV}^{-1}$, $g_{13} = g_{ae} / 10^{-13}$, $C^{B/C/P}$ are the normalizations, $E_0^{B/C/P}$ are the cut-off energies, and $\beta^{B/C/P}$ are the spectral indices for bremsstrahlung, Compton, and Primakoff processes, respectively.
    This follows the description of \citet{xiao2022betelgeuse} which uses the Full Network Stellar evolution code (FuNS~\citep{straniero2019initial}) to derive the fluxes and spectral shapes.
	The values for these parameters in the case of Betelgeuse had been obtained from previous simulations~\citep{xiao2022betelgeuse} for different times to core-collapse ($\mathrm{t_{cc}}$).
    The fluxes depend on the temperature and density conditions directly derived from the hydrodynamics profiles provided by the code.
    All the stellar models considered lead to a surface luminosity able to reproduce the position of these stars in the Hertzsprung-Russell diagram.
    The models cover a wide range of stellar evolutionary phases which reproduce the observational data.
	The closer a star is to core-collapse, the hotter is its core, and therefore should have a higher ALP production rate, 
  resulting in a higher $\gamma$-ray flux.
    In Fig.~\ref{fig:ALP_spectrum}, we show the expected ALP flux contributions from the three production mechanisms as well as their combined spectrum for Betelgeuse at a time to core-collapse of $6900$~yr. 
    For comparison, we indicate the energy upper limit of NuSTAR ($79$~keV; blue dashed line), highlighting that the bulk of the ALP-induced emission lies above NuSTAR's sensitivity and extends into the soft $\gamma$-ray regime. 
    This emphasizes the suitability of using INTEGRAL/SPI, to probe this energy range. \\
The differential photon flux per unit energy arriving at Earth is
	\begin{equation}\label{eq:ALP_differntial_flux}
		\dfrac{dN_\gamma}{dE\,dS\,dt} = \dfrac{1}{4\pi d^2}\dfrac{d\dot{N}_a}{dE}P_{a\gamma} \equiv \mathrm{F}(E_{\mathrm{ini}}, m_a, g_{a\gamma}, g_{ae})\mathrm{,}
	\end{equation}
	where $P_{a\gamma}$ is the ALP-photon conversion probability given by
	\begin{equation}
		P_{a\gamma} = 8.7 \times 10^{-6}g^2_{11}\Bigg(\dfrac{B_T}{1\mu\mathrm{G}}\Bigg)^2\Bigg(\dfrac{d}{197~\mathrm{pc}}\Bigg)^2\dfrac{\sin^2(qd)}{(qd)^2}\mathrm{.}
		\label{eq:ALP_conversion_probability}
	\end{equation}
	Here, $B_T$ is the Galactic magnetic field, $d$ is the distance to the star, and $q$ is the momentum transfer~\citep{xiao2022betelgeuse}. 
    The product of the momentum transfer $q$, and the distance $d$ is given by
    \begin{equation}
    \resizebox{0.92\linewidth}{!}{$
        qd \simeq \left[77 \left(\dfrac{m_a}{10^{-10}~\mathrm{eV}}\right)^2 -0.14\left(\dfrac{n_e}{0.013~\mathrm{cm}^{-3}}\right)\right] \times \left(\dfrac{d}{197~\mathrm{pc}}\right) \left(\dfrac{E}{1~\mathrm{keV}}\right)^{-1} 
        $}\mathrm{.}
    \end{equation}
    Here, $m_a$ is the mass of the ALP and $n_e$ is the electron density.
    For sources within $<1$~kpc, we assume a uniform electron density ($n_e = 0.013~\mathrm{cm}^{-3}$) and three Galactic magnetic field scenarios with uniform Galactic magnetic field ($B_T = 0.4, 1.4, 3.0~\mu\mathrm{G}$).
    These values are motivated by measurements of Betelgeuse as also quoted in~\citep{xiao2022betelgeuse}.
    Since the Galactic magnetic field at $\sim$hundred pc distances is not very well modeled, we use Betelgeuse as indicative, and using these three above mentioned $B_T$ scenarios help us cover a wide range of realistic possibilities.
    The given spectrum and expected flux contributions are specifically created for Betelgeuse.
    However, we also use it for other stars even though the parameters might depend on the mass of the star.
    The mass distribution of our stellar sample ranges from $5~\mathrm{M}_{\odot} - 29~\mathrm{M}_{\odot}$ with the average mass of a star being $\sim 11.5~\mathrm{M}_{\odot}$.
    The average star mass is comparable to Betelgeuse, and due to our combined analysis approach, it is a fair assumption to apply this same spectrum to other stars.

\section{Dataset and analysis}\label{sec:dataset_analysis}
\subsection{SPI data analysis}\label{subsec:SPI}
	We use INTEGRAL/SPI observations in the energy range $20 - 2000$\,keV, with a field of view of $20^\circ$, around the regions of Orion/Betelgeuse ($4.4~\mathrm{Ms}$, 1770 pointings, 2486 sec per pointing), Cygnus ($18.3~\mathrm{Ms}$, 7463 pointings, 2452 sec per pointing), Pegasus ($1.1~\mathrm{Ms}$, 482 pointings, 2282 sec per pointing), and Spica ($0.5~\mathrm{Ms}$, 239 pointings, 2092 sec per pointing).
	We select pointings that fall within a $10^\circ$ radius around our source of interest namely Betelgeuse (l = $199.787^\circ$, b = $-8.959^\circ$), Rigel (l = $209.24^\circ$, b = $-25.24^\circ$), Spica (l = $-43.88^\circ$, b = $50.84^\circ$), Pegasus (l = $76^\circ$, b = $-37^\circ$), and for Cygnus a rectangular region spanning (l = [$100^\circ$, $130^\circ$], b = [$-10^\circ$, $10^\circ$])  (shown in Fig.\ref{fig:Multiple_sources} with exposure outlines).
    Data from SPI are modeled in the following way:
    \begin{equation}
        d_{i,j,k} = \sum_{l} R_{l; ijk} \sum_{n=1}^{N_\mathrm{s}} \alpha_{nk} S_{nl} + \sum_{n=N_\mathrm{s}+1}^{N_\mathrm{s}+N_\mathrm{b}} \beta_{nk} B_{n;ijk}\mathrm{,}
        \label{eq:SPI_data_modeling}
    \end{equation}
    where, $d_{i,j,k}$ is the event counts with $i, j, k$ being the indices of pointing, detector, and energy bin that spans the data space, respectively.
    $R_{l; ijk}$ is the instrument response for a given sky direction, $l$.
    This also includes the effective area and determines the point spread function.
    The energy dispersion is included in the spectral fit which are named as $E_\mathrm{ini}, E_\mathrm{fin}$.
    $\alpha_{nk}$ and $\beta_{nk}$ are the normalization factors (model parameters) for the $N_\mathrm{s}$ sky model components, $S_{n}$, and $N_\mathrm{b}$ background model components, $B_{n}$, respectively~\citep{diehl2018integral}. 
    We assume no a priori spectra, so the extracted data points for each source (flux values), correspond to $\alpha_{nk}$ for each of the $n$ sources.
	The background is created from the SPI background and response data base \citep{diehl2018integral, siegert2019background}. 
    The background at a specific energy bin is modeled using two components: photons from continuum processes, and photons from $\gamma$-ray lines.
    For a specific physical process inside the satellite, the detector patterns from the background stay constant.
    The only thing that might change as a function of time/pointing is the amplitude of the two background components which is determined in a maximum likelihood fit.
	The sky model is defined by a list of already known SPI point sources~\citep{bouchet2008integral} (Fig.~\ref{fig:Multiple_sources} orange), the diffuse positron annihilation signal, and the diffuse 1.8\,MeV line from the decay of ~\nuc{Al}{26}  and any additional source one would like to fit.
    In the Betelgeuse dataset, this would mean 5 SPI point sources, plus our 4 red super giant candidates (Betelgeuse, Rigel, CE Tauri, and S Monocerotis A at their respective positions), plus the diffuse positron annihilation map \citep{Siegert2016_511} for 467.5--514\,keV, plus the diffuse \nuc{Al}{26} map for the 1809\,keV bin.
	To account for any possible strong and/or variable source contamination such as from the Crab, narrow energy bins (0.5\,keV bins from 20 -- 105\,keV, 2\,keV bins from 105 -- 203\,keV, and logarithmic bins from 203 -- 2000\,keV) were used for the Betelgeuse region.
    Such narrow energy bins would result in oversampling the energy dispersion.
    The oversampling is only to account for the highly variable background (strong lines) and is not relevant here because the ALP model spectra are broadband.
    For the other three regions, logarithmic energy bins were found sufficient and narrow bins did not significantly affect the quality of the fit.
    Finally, we use OSA/spimodfit \citep{Courvoisier03, Strong2005_gammaconti} to extract the flux per energy bin by fitting the SPI data using Eq.\,(\ref{eq:SPI_data_modeling}).
	After an initial maximum likelihood fit to extract the SPI spectrum from the raw detector counts by fitting the sky and background model components independently in each energy bin Eq.~\ref{eq:SPI_data_modeling}, we found that some of the observations (Betelgeuse region: 8\%, Cygnus region: 3\%) were contaminated by investigating the residuals as a function of pointing ID. 
    We iteratively removed these ``bad pointings'' by clipping outliers which deviate from our expectation by more than $5\sigma$. 
    Such outliers typically originate in solar particle events or transients which are not modeled in this approach. 
    This results in typical goodness-of-fit values of $\chi^2/\mathrm{dof}$ of Orion: $1.02 \pm 0.02$, Cygnus: $0.89 \pm 0.02$,  Pegasus: $0.99 \pm 0.03$, Spica: $1.00 \pm 0.02$, which is adequate given the number of data points in each dataset being around 33630, 141797, 9158, and 4541 respectively.
    We quote the mean and standard deviation of these per bin $\chi^2/\mathrm{dof}$ values to demonstrate the quality of the background and sky modeling in each region.
    These values only serve as a diagnostic of SPI spectral extraction.

\begin{figure}[ht!]
    \centering
    \includegraphics[width=\linewidth]{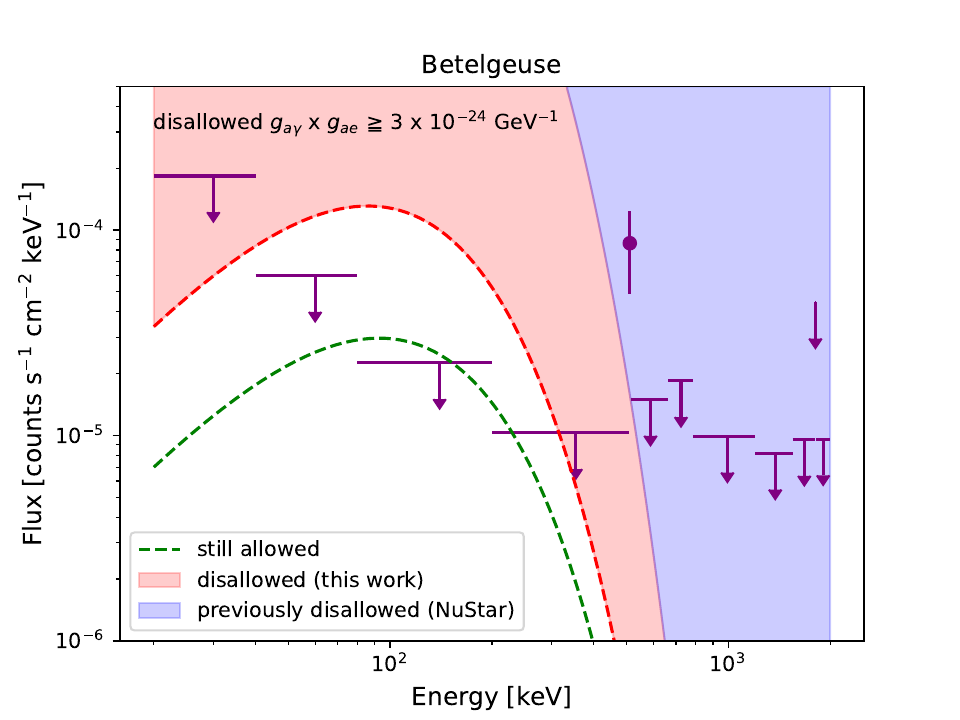}
    \caption{Betelgeuse spectrum as obtained from SPI for the energy range $20$ -- $2000$~keV. The dot with the error bar shows the flux value in that energy bin. The downward arrows show the $3\sigma$ upper limit for bins where the flux significance is less than $2\sigma$. The 511 keV bin is systematically large because of incomplete modelling of the diffuse emission in the Crab/Orion region and is therefore not taken as a detection of 511 keV in Betelgeuse. The red shaded region shows the flux from 
 the $g_{a\gamma} \times g_{ae}$ values allowed by NuSTAR that can now directly be excluded from the ALP parameter space since the flux prediction from them is larger than the $3\-\sigma$ flux limits from SPI. The excluded limit is $g_{a\gamma} \times g_{ae} \geqq 3\times 10^{-24}~\mathrm{GeV}^{-1}$. The blue shaded region shows the flux from the $g_{a\gamma} \times g_{ae}$ values that were already disallowed in the NuSTAR study~\citep{xiao2022betelgeuse}.}
		\label{fig:Bet_spectra}
\end{figure}

	The $\gamma$-ray spectrum obtained from SPI for Betelgeuse in particular, and all other 17 sources as well, is consistent  with zero, leading us to estimate upper limits for the two couplings.
	Fig.~\ref{fig:Bet_spectra} shows the extracted SPI spectrum of Betelgeuse to which different ALP spectra (age:  $1.55 \times 10^5$--$3.6$\,yr, Galactic magnetic field $B_T$: $0.4, 1.4, 3.0~\mu\mathrm{G}$, distance $d$: $222$\,pc, electron density $n_e$: $0.013~\mathrm{cm}^{-3}$, and ALP mass $m_a$: $10^{-14}$--$10^{-9}$\,eV) were fitted.
   The spectra are fitted with a Gaussian likelihood, corresponding to $\chi^2$,
    \begin{equation}
    \resizebox{0.92\linewidth}{!}{$
        \chi^2(\alpha_k | \mathrm{M}_k(m_a, g_{a\gamma}, g_{ae})) = \sum\limits_{k} \left(\dfrac{\alpha_k - \mathrm{M}_k(m_a, g_{a\gamma}, g_{ae})}{\sigma_{\alpha_k}} \right)^2
         $}
        \label{eq:chi_squared}
    \end{equation}
    with 
    \begin{equation}
    \resizebox{0.92\linewidth}{!}{$
        \mathrm{M}_k(m_a, g_{a\gamma}, g_{ae}) = \dfrac{1}{\Delta E_k} \int_{E_{\mathrm{min},k}}^{E_{\mathrm{max},k}} \mathrm{F(E_{\mathrm{ini}}, m_a, g_{a\gamma}, g_{ae})} \cdot \mathrm{R}(E_{\mathrm{ini}}) \mathrm{d}E_{\mathrm{ini}}
        $}\mathrm{,}
        \label{eq:model_spectrum}
    \end{equation}
    where $\alpha_k$ is the extracted flux values, $\sigma_{\alpha_k}$ are the uncertainty on the flux values, and $\mathrm{R}$ is the instrument response matrix.
    This $\chi^2$ is the statistical quantity that is used to evaluate the ALP model parameters and is unrelated to the $\chi^2/\mathrm{dof}$ values quoted for the SPI spectral extraction process.
    Since the spectrum is consistent with zero, we show the $3\sigma$ upper limits for energy bins where the flux significance is $< 2\sigma$.
    In the red shaded region, we also show the flux allowed by earlier constraints, such as from the NuSTAR study by~\citet{xiao2022betelgeuse}, which is now disallowed in this study since the flux expectation is higher than the $3\sigma$ flux measured by SPI.
	Similarly, the blue shaded region are the allowed flux values that were already rejected in the NuSTAR study.
	For the spectral analysis, we use 14 different ALP spectra combinations, 12 of which are for $\mathrm{B_T} = 1.4~\mu\mathrm{G}$ and $\mathrm{t_{cc}}$ ranging from $1.55 \times 10^5$--$3.6$\,yr~\citep{xiao2022betelgeuse}, one for $\mathrm{B_T} = 3.0~\mu\mathrm{G}$ and $\mathrm{t_{cc}} = 3.6~\mathrm{yr}$, and one for $\mathrm{B_T} = 0.4~\mu\mathrm{G}$ and $\mathrm{t_{cc}} = 1.55\times10^5~\mathrm{yr}$.
    Such a wide range of spectral model combinations ensure that the entire range of possibilities is accounted for, ranging from very conservative case (low Galactic magnetic field and early phases of stellar evolution) to most optimistic case (high Galactic magnetic field and late stages of stellar evolution).
    Therefore, any possible realistic scenario would always lie in the range of coupling constants obtained from these two extreme cases.

\subsection{Hierarchical modelling approach}\label{subsec: BHM}
\begin{figure}[ht!]
    \centering
    \includegraphics[width=\linewidth]{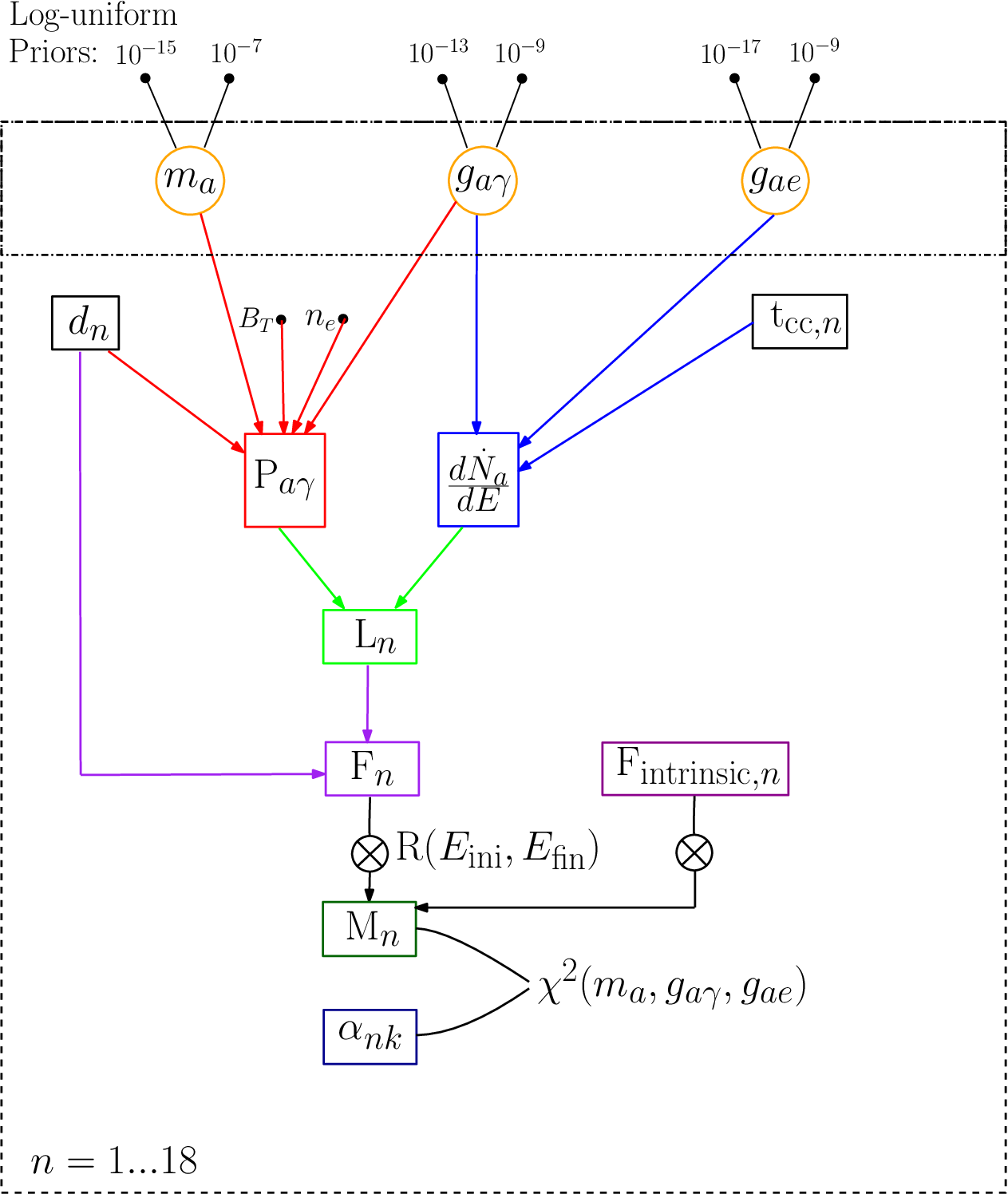}
    \caption{Bayesian hierarchical model used to constrain the ALP parameters $m_a$, $g_{a\gamma}$, $g_{ae}$. The model assumes that these global parameters are shared across all 18 sources and govern both ALP production in stellar interiors and their conversion to photons in the Galactic magnetic field. Each star contributes a predicted photon flux  based on its luminosity, distance, and a shared ALP spectral shape. This flux is convolved with the instrument response to yield the expected counts which are compared to the observed data. A joint likelihood analysis is performed across all sources using the 3ML framework to obtain constraints on the ALP parameter space.}
    \label{fig:BHM_axion}
\end{figure}

Fig.\ref{fig:BHM_axion} presents the Bayesian hierarchical model used to jointly constrain the properties of ALPs from $\gamma$-ray observations of the 18 sources. 
The three key ALP parameters, the ALP mass $m_a$, ALP-photon coupling $g_{a\gamma}$, and the ALP-electron coupling $g_{ae}$, are treated as global parameters, linked across all sources, and were assigned priors in a Bayesian inference framework. 
These govern both the production of ALPs in stellar interiors and their conversion to photons in the Galactic magnetic field. 
To perform the combined analysis, we assume a shared spectral shape for the ALP induced $\gamma$-ray flux across all stars ($n= 1...18$) as the underlying physical mechanisms in all stars remain the same.
This spectral shape is derived from standard ALP emission processes (Compton, Primakoff, bremsstrahlung) and depends on $\mathrm{t}_{\mathrm{cc}}$,  $g_{a\gamma}$, $g_{ae}$.
The ALP-photon conversion probability $\mathrm{P}_{a\gamma} (m_a, g_{a\gamma}, B_T, n_e, d_n)$ further modifies this spectrum, depending on the fixed parameters $B_T = (0.4, 1.4, 3.0)~\mu\mathrm{G}$, $n_e = 0.013~\mathrm{cm}^{-3}$, and the distance to each star $d_n$, which may also be uncertain but our assumption of a uniform magnetic field helps us eliminate the distance dependence in the expected photon flux.
While the spectral normalization varies from star to star due to differences in $d_n$ (scaling as $1/4\pi d_n^{-2}$), the underlying spectral shape and the governing ALP parameters are common to all sources.
Each star’s predicted photon flux $F_n$ is obtained by scaling the ALP luminosity, $L_n = {d\dot{N}_a}/{dE}\, \times P_{a\gamma}$ by the respective distance (see Eq.\,\ref{eq:ALP_differntial_flux}).
The smooth astrophysical models in units of $\mathrm{ph\,cm^{-2}\,s^{-1}\,keV^{-1}}$ are then converted to observable counts by applying the instrument response matrix $\mathrm{R}(E_\mathrm{ini}, E_\mathrm{fin})$ (Eq.~\ref{eq:model_spectrum}). 
An additional component $F_{\mathrm{intrinsic}, n}$ may account for known $\gamma$-ray line or continuum emissions, including the $511$~keV positron annihilation line and positronium continuum, and the $1.809$~MeV line from \nuc{Al}{26} decay unless modeled already in the spectral extraction step.
%
%
The full model is fitted to the observed data $\alpha_{nk}$ for each star via a likelihood function (Eq.~\ref{eq:chi_squared}), and the fit is performed jointly across all 18 stars using the 3ML (Multi-Mission Maximum Likelihood) framework \citep{vianello20153ML}.
3ML allows for shared parameters across multiple datasets, enabling a coherent global fit in which the particle physics parameters $m_a$, $g_{a\gamma}$, $g_{ae}$ are constrained simultaneously using all available information and uncertainties.
The model thus fully exploits the consistency of ALP physics across stellar environments while accounting for differences in source distances, instrumental responses, and intrinsic background features.

%
	\begin{figure}[h!]
		\includegraphics[width=\linewidth]{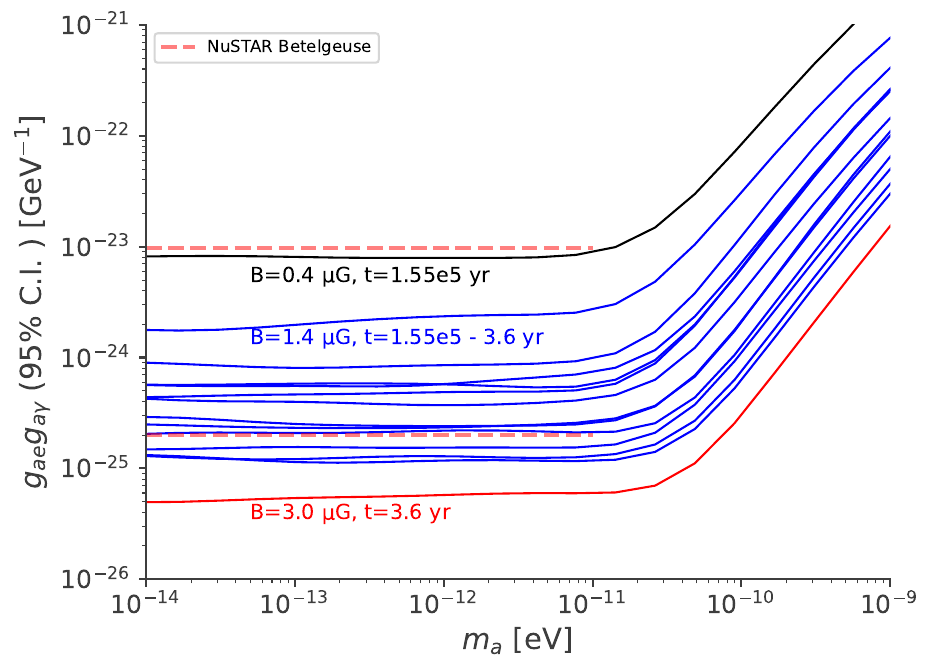}
		\centering
		\caption{The 95\% C.I. upper limits of $g_{ae} \times g_{a\gamma}$ as a function of ALP mass for Betelgeuse in the energy range $20 - 600$\,keV. The solid blue lines show the upper limit for each stellar model, assuming a representative value of $B_T = 1.4~\mu\mathrm{G}$ with the top blue line corresponding to $\mathrm{t}_{\mathrm{cc}}=1.55\times10^5~\mathrm{yr}$ and the bottom blue line corresponding to $\mathrm{t}_{\mathrm{cc}}=3.6~\mathrm{yr}$. The constraints will scale with different $B_T$ as in Eq.~\ref{eq:ALP_conversion_probability}, the solid black line shows the upper limit for the most conservative ($B_T = 0.4~\mu\mathrm{G}$ and $t_{cc} = 1.55 \times 10^5 \mathrm{yrs}$) and the solid red line for the most optimistic case ($B_T = 3.0~\mu\mathrm{G}$ and $t_{cc} = 3.6 \mathrm{yrs}$). For comparison, we also show the 95\% C.I. upper limits of $g_{ae} \times g_{a\gamma}$ obtained from \citet{xiao2022betelgeuse} for the most conservative and optimistic cases with dashed red lines.}
		\label{fig:Bet_constraints}
	\end{figure}
	%

\section{Results}\label{sec:results}
	\subsection{Individual Stars}\label{subsec:indiv_stars}
    
    We use a representative ALP spectrum \citep[also taken from][]{xiao2022betelgeuse} to fit our source spectrum and estimate the parameter values of $g_{ae} \times g_{a\gamma}$ for a given range of $m_a$. 
	Fig.~\ref{fig:Bet_constraints} shows the 95\% C.I. obtained from Betelgeuse for our set of available stellar models.
    We find that the constraints obtained from SPI for Betelgeuse are a factor of $\sim5$ better than those estimated from NuSTAR.
    This improvement is also shown in Fig.~\ref{fig:Bet_constraints}.

\begin{figure}[ht]
		\includegraphics[width=\linewidth]{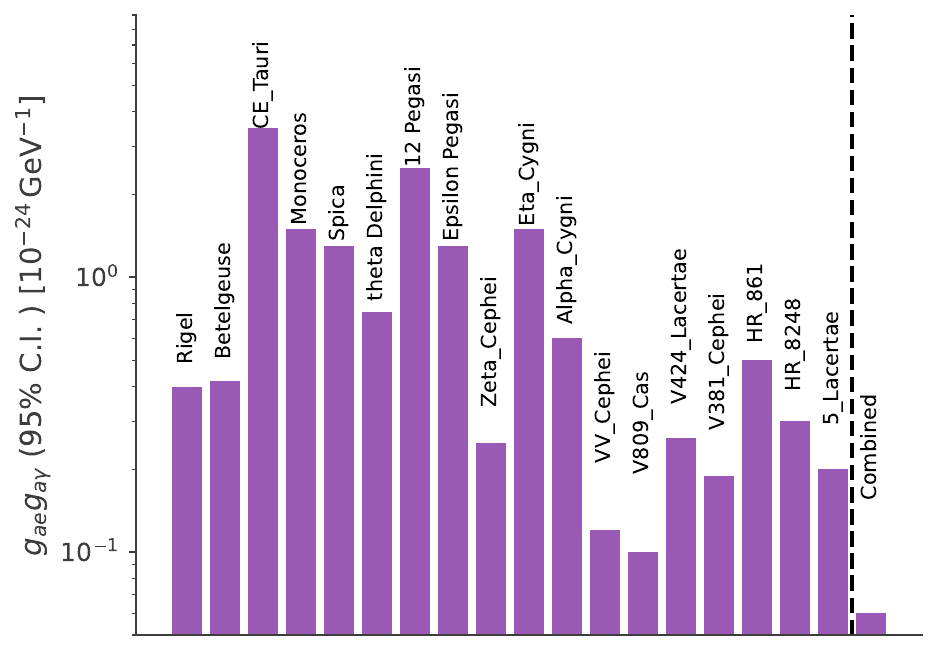}
		\centering
		\caption{The 95\% C.I. upper limits of $g_{ae} \times g_{a\gamma}$ for the entire star sample (individual stars and from a combined analysis) for $B_T = 1.4~\mu\mathrm{G}$ and $t_{cc} = 6900 \mathrm{yrs}$ and $m_a < 10^{-11}$~eV.}
		\label{fig:all_stars_bar}
	\end{figure}

    In Fig.~\ref{fig:all_stars_bar}, we show the 95\% C.I. obtained for each individual star and from a combined analysis of all stars (see \ref{subsec: BHM}) for a representative stellar model $\mathrm{t}_{\mathrm{cc}}=6900~\mathrm{yr}$ and $B_T = 1.4~\mu\mathrm{G}$.
	%
	We find that not all stars result in the same limits, as expected from the different exposure times and distances.
	We assume each star can be modeled by the spectrum estimated for Betelgeuse \citep{xiao2022betelgeuse}.
    While the detailed spectral shape may vary with stellar mass, the underlying physical processes governing ALP production are expected to be similar for stars in comparable evolutionary stages.
    Additionally, we assume that all stars are in the same burning phase.
    These assumptions ensure that any realistic scenario would lie within the range spanned by the various combinations of this model, thereby providing a range of possible constraints.
    We estimate that using this single stellar evolution model for a $12\,\mathrm{M_\odot}$ star like Betelgeuse will result in a systematic uncertainty.
    Due to the stiff dependence on the core temperature, typical uncertainties on stellar profiles may lead to uncertainties which can be maximally estimated within $\sim1$ order of magnitude of difference in the fluxes, leading to uncertainties within a factor $\sim2-3$ on the resulting limits.
    This same mass assumption and uncertainty is negligible compared to the uncertainties on the time to core-collapse or the magnetic field on the line-of-sight between the detector and the respective star.
	Additionally, some regions in the sky are observed more often than others, giving us better statistics for the respective sources such as for the sources in the Cygnus region. 
	Nevertheless, as reported by \citet{xiao2022betelgeuse}, their analysis can be extended essentially unchanged to other close-by supergiant stars strengthening the credibility of this analysis. 

\begin{table*}
    \centering
    \renewcommand{\arraystretch}{1.5} 
    \begin{tabularx}{\linewidth}{lrrccrcrrr}
    \toprule
    \toprule
    Common Name & Gal. long & Gal. lat & Distance & Mass & $g_{ae} \times g_{a\gamma}$  & $g_{a\gamma}$ & $511~\mathrm{keV}$ & $1.8~\mathrm{MeV}$ & $M_{\mathrm{Al}}$ [$\mathrm{M_{\odot}}$] \\
    & [deg] & [deg] & [kpc] & [$\mathrm{M_{\odot}}$] & $2\sigma$~UL & $2\sigma$~UL & $3\sigma$~UL & $3\sigma$~UL & $3\sigma$~UL \\
    \midrule 

    Spica & 316.11 & 50.84 & $0.077 \pm 0.004$ & $11.43^{+1.15}_{-1.15}$ & $27.7$ & 4.36 & $35.6$ & $11.8$ & 6 \\

    $\epsilon$ Pegasi & 65.57 & -31.45 & $0.211 \pm 0.006$ & $11.7^{+0.8}_{-0.8}$ & $13.5$ & 3.26 & $30.3$ & $9.7$ & 37 \\

    Betelgeuse & 199.78 & -8.95 & $0.222 \pm 0.040$ & $11.6^{+5.0}_{-3.9}$ & $8.1$ & 2.15 & $11.1$ & $4.4$ & 19 \\

    $\zeta$ Cephei & 103.06 & 1.66 & $0.256 \pm 0.006$ & $10.1^{+0.1}_{-0.1}$ & $3.3$ & 1.97 & $1.1$ & 0.4 & 2 \\
    
    Rigel & 209.24 & -25.24 & $0.264 \pm 0.024$ & $21.0^{+3.0}_{-3.0}$ & $11.7$ & 3.31 & $5.7$ & $1.8$ & 11 \\
   
    $\xi$ Cygni & 71.01 & 3.36 & $0.278 \pm 0.029$ & $8.0$ & $25.1$ & 5.69 & $13.4$ & $4.6$ & 30 \\

    S Monocerotis A & 202.93 & 2.19 & $0.282 \pm 0.040$ & $29.1$ & $41.8$ & 5.51 & $45.9$ & 13.3 & 90 \\

    CE Tauri & 187.17 & -8.07 & $0.326 \pm 0.070$ & $14.37^{+2.00}_{-2.00}$ & $50.8$ & 5.86 & $289.0$ & $41.4$ & 375 \\
    
    12 Pegasi & 76.64 & -22.82 & $0.415 \pm 0.031$ & $6.3^{+0.7}_{-0.7}$ & $56.4$ & 6.68 & $100.0$ & $17.0$ & 250 \\
    
    5 Lacertae & 99.66 & -8.65 & $0.505 \pm 0.046$ & $5.11^{+0.18}_{-0.18}$ & $17.2$ & 3.60 & $1.9$ & $0.7$ & 15 \\

    VV Cephei & 104.92 & 7.04 & $0.599 \pm 0.083$ & $10.6^{+1.0}_{-1.0}$ & $4.7$ & 3.11 & $1.1$ & $0.4$ & 13 \\
    
    $\theta$ Delphini & 57.94 & -16.60 & $0.629 \pm 0.029$ & $5.60^{+3.0}_{-3.0}$ & $22.5$ & 4.98 & $45.8$ & $14.9$ & 503 \\
    
    V381 Cephei & 98.18 & 6.35 & $0.631 \pm 0.086$ & $12.0$ & $5.1$ & 2.01 & $2.3$ & $0.9$ & 31 \\
    
    V424 Lacertae & 104.58 & -8.99 & $0.634 \pm 0.075$ & $6.8^{+1.0}_{-1.0}$ & $4.2$ & 4.93 & $1.2$ & $0.5$ & 16 \\
    
    HR 861 & 135.95 & 4.66 & $0.639 \pm 0.039$ & $9.2^{+0.5}_{-0.5}$ & $14.4$ & 2.30 & $1.9$ & $0.7$ & 25 \\
    
    V809 Cassiopeia & 112.61 & 1.72 & $0.730 \pm 0.074$ & $8.3^{+0.5}_{-0.5}$ & $2.0$ & 1.19 & $0.7$ & $0.3$ & 13 \\
    
    HR 8248 & 90.83 & -4.30 & $0.746 \pm 0.039$ & $6.3^{+0.7}_{-0.7}$ & $4.6$ & 1.66 & $4.7$ & $1.8$ & 85 \\
    
    $\alpha$ Cygni & 84.28 & 1.99 & $0.802 \pm 0.066$ & $19.0^{+4.0}_{-4.0}$ & $11.5$ & 3.58 & $11.5$ & $3.9$ & 213 \\
   
    \hline 
    combined: & & & & & $1.9$ & $1.26$ & & & \\
    \bottomrule
    \end{tabularx}
    \caption{List of massive stars used in this analysis along with their Galactic coordinates, distance and mass (obtained from~\citet{mukhopadhyay2020presupernova}). We also show the most conservative ($B_T = 0.4~\mu\mathrm{G}, \mathrm{t_{cc}} =  1.55\times10^5~\mathrm{yr}$) 95\% confidence interval upper limit obtained for the coupling product $g_{ae} \times g_{a\gamma}$ and for the photon only coupling $g_{a\gamma}$. In addition, the $3\sigma$ flux upper limits for the 511\,keV and the 1809\,keV are also listed. Finally, the \nuc{Al}{26} mass expected based on the 1809\,keV flux is also shown. The units of $g_{ae} \times g_{a\gamma}$, $g_{a\gamma}$, the line fluxes, and the \nuc{Al}{26} masses are $10^{-24}\,\mathrm{GeV^{-1}}$, $10^{-11}\,\mathrm{GeV^{-1}}$, $10^{-5}\,\mathrm{ph\,cm^{-2}\,s^{-1}}$, and $10^{-5}\,\mathrm{M_\odot}$, respectively.}
    \label{tab:preSN_sources}
\end{table*}

\begin{figure}[ht!]
	\includegraphics[width=\linewidth]{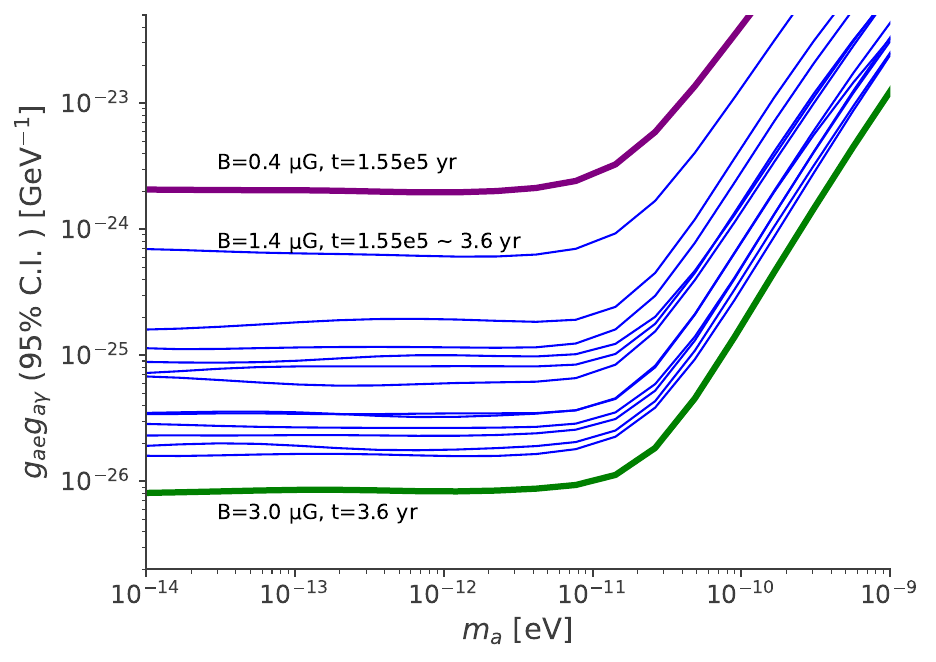}
	\centering
	\caption{The 95\% C.I. upper limits of $g_{ae} \times g_{a\gamma}$ as a function of ALP mass $m_a$ for the combined analysis of all 18 stars for the energy range $20 - 600~\mathrm{keV}$. The solid blue lines show the upper limit for each stellar model, assuming a representative value of $B_T = 1.4~\mu\mathrm{G}$ with the top blue line corresponding to $\mathrm{t}_{\mathrm{cc}}=1.55\times10^5~\mathrm{yr}$ and the bottom blue line corresponding to $\mathrm{t}_{\mathrm{cc}}=3.6~\mathrm{yr}$. The solid purple line shows the upper limit for the most conservative ($B_T = 0.4~\mu\mathrm{G}$ and $t_{cc} = 1.55 \times 10^5~\mathrm{yr}$) and the solid green line for the most optimistic case ($B_T = 3.0~\mu\mathrm{G}$ and $t_{cc} = 3.6~\mathrm{yr}$).}
	\label{fig:all_stars_limits}
    \end{figure}

\begin{figure}[ht!]
        \includegraphics[width=\linewidth]{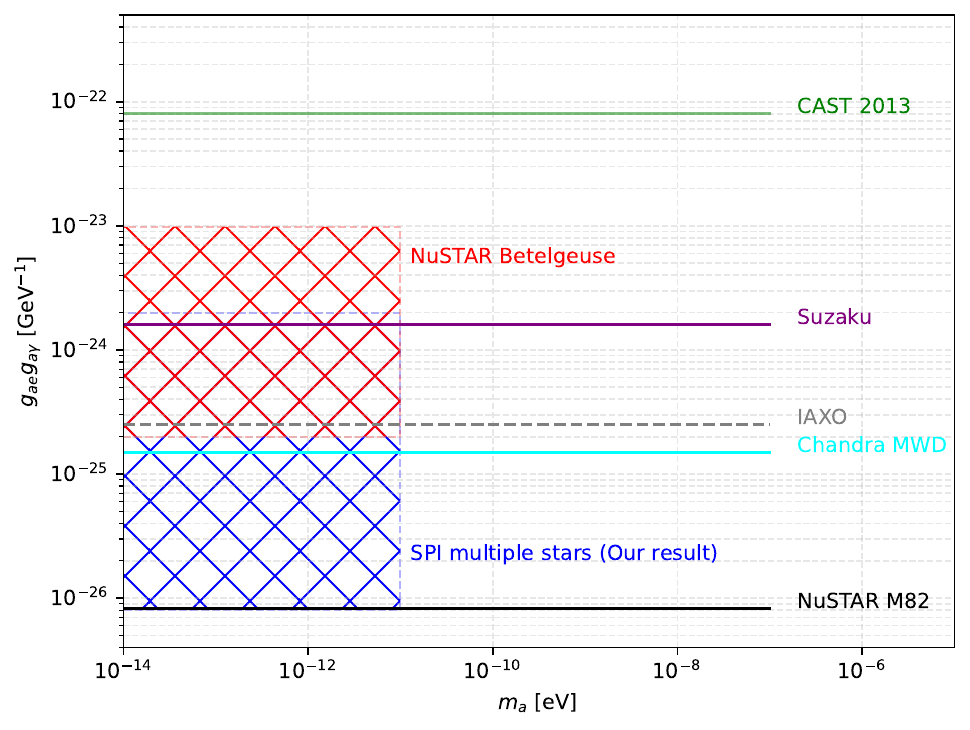}
        \centering
        \caption{Comparison of $g_{ae} \times g_{a\gamma}$ across different instruments and different astrophysical objects. The bounds for CAST are obtained from~\citet{barth2013cast}, NuSTAR's Betelgeuse bounds from~\citet{xiao2022betelgeuse}, Suzaku bounds from~\citet{dessert2019Suzaku}, projected sensitivity of IAXO from~\citet{abeln2021IAXO}, Chandra's MWD study from~\citet{dessert2022ChandraMWD}, and NuSTAR's M82 bounds from~\citet{Ning:2025tit}. This study improves the previous limits by over an order of magnitude for the most optimistic case. The study of M82 with NuSTAR might still provide the tightest constraints in the literature, however the limitation of their analysis is discussed in the conclusion.}
        \label{fig:comparison_multi_instruments}
    \end{figure}

	Fig.~\ref{fig:all_stars_limits} shows the 95\% C.I. obtained from a combined analysis of all 18 stars for all stellar models ($\mathrm{t_{cc}}$ = $1.55 \times 10^5$--$3.6$\,yr) for the energy range $20-600~\mathrm{keV}$ (extension to higher energies is shown in the Appendix~\ref{sec:appendix}).
    We find the combined coupling range to be $g_{a\gamma} \times g_{ae}$ = ($0.008$ -- $2$) $\times 10^{-24}~\mathrm{GeV}^{-1}$.
    This result improves the previous estimation by $\sim 1$ order of magnitude for the most conservative stellar model ($\mathrm{t_{cc}} = 1.55\times10^5$~yr, $B_T = 0.4~\mu\mathrm{G}$) and about a factor of 25 for the most optimistic (but unrealistic, see next section) case ($\mathrm{t_{cc}} = 3.6$~yr, $B_T = 3.0~\mu\mathrm{G}$).
    We also place limits on the ALP-photon coupling, $g_{a\gamma}$, by assuming Primakoff emission as the only viable production channel and in which the ALP-electron coupling is switched off.
    We find the 95\% C.I. on $g_{a\gamma}$ to be in the range ($0.13$ -- $1.26$) $\times 10^{-11}~\mathrm{GeV}^{-1}$ depending on the magnetic field model and the time to core-collapse.
    This is a small improvement on the previous limits by \citet{Xiao:2020pra} which sets the 95\% C.I. on the ALP-photon coupling of $g_{a\gamma} < (0.5 - 1.8)\times10^{-11}~\mathrm{GeV}^{-1}$ .
    The range of parameters obtained from this study of multiple stars with SPI also improves on estimations from previous studies performed with different instruments and alternative candidate sources \citep[e.g.,][and references therein]{barth2013cast, dessert2019Suzaku, abeln2021IAXO, dessert2022ChandraMWD}.
	This comparison is shown in Fig.~\ref{fig:comparison_multi_instruments}.
    \citet{Ning:2025tit} sets stronger constraints on the coupling product than our analysis, however the limitation of their study is discussed in the conclusion (Sec.~\ref{sec:conclusion}).

\subsection{Conservative results}\label{subsec:likely_results}

\begin{figure}[ht]
        \includegraphics[width=\linewidth]{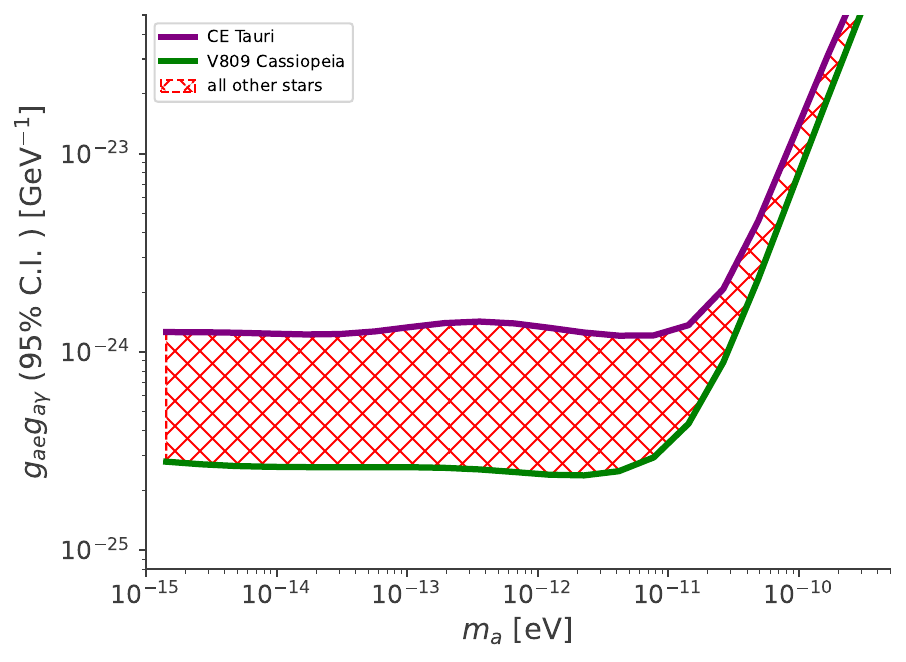}
        \centering
        \caption{Range of 95\% C.I upper limits on $g_{a\gamma} \times g_{ae}$ as a function of ALP mass $m_a$, derived from the conservative scenario in which one star out of 18 is assumed to be in a later evolutionary stage ($\mathrm{t}_{\mathrm{cc}}=2.3\times10^4$~yr), while the remaining 17 stars are fixed at an earlier He-burning phase ($\mathrm{t}_{\mathrm{cc}}=1.55\times10^5$~yr) and a uniform $B_T=0.4~\mu\mathrm{G}$. The strongest constraint is obtained when V809 Cassiopeia is assumed to be closer to core collapse (green), and the weakest when CE Tauri is (purple). The hatched region is the range of constraints obtained when every other star is individually placed in the more advanced He-burning phase. This method accounts for uncertainties in stellar evolution and provides a realistic range of possible limits.}
        \label{fig:conservative_scenario_range}
    \end{figure}

To obtain more conservative and realistic constraints on ALP couplings, we consider a scenario in which all but one star in our sample is assumed to be in an early stage of stellar evolution. 
 This is specifically the early Helium burning phase corresponding to $\mathrm{t}_{\mathrm{cc}}=1.55\times10^5$~yr.
One star at a time is then placed in a more advanced phase with time to core-collapse of $\mathrm{t}_{\mathrm{cc}}=2.3\times10^4$~yr, representing a later stage of Helium burning.
For this conservative scenario, we assume a uniform Galactic magnetic field of $B_T = 0.4~\mu\mathrm{G}$.
This approach avoids the overly optimistic assumption that all stars are simultaneously near core collapse and allows us to explore the effect of evolutionary differences on the derived limits. 
By iteratively assigning the more advanced stage to each of the 18 stars, while keeping the others in the early phase, we obtain a range of constraints that reflect the potential diversity in the actual stellar states.
This method also mitigates the variability seen in individual star constraints (as illustrated in Fig.~\ref{fig:all_stars_bar}), providing a more balanced estimate of the overall sensitivity.
Fig.~\ref{fig:conservative_scenario_range} shows the resulting range of 95\% C.I upper limits on the product $g_{a\gamma} \times g_{ae} = (0.27 - 1.25) \times 10^{-24}~\mathrm{GeV}^{-1}$.
The strongest constraint is obtained when V809 Cassiopeia is placed in the advanced stage ($\mathrm{t}_{\mathrm{cc}}=2.3\times10^4$~yr), while the weakest arises when CE Tauri is assumed to be the star closer to collapse.
This outcome is consistent with the individual star sensitivities shown earlier and emphasizes how the depth of observation in each region influences the results.
This conservative framework provides a more robust and physically motivated estimate of the ALP parameter space, accounting for astrophysical uncertainties in stellar evolution stages.
%

	\subsection{Gamma-ray lines}\label{subsec:gamma-ray_lines}
While the focus of this work is to search for ALPs, we also investigate the possibility of these red super giants to show significant $\gamma$-ray line emission at 511\,keV (positron annihilation) and 1809\,keV (decay of radioactive \nuc{Al}{26}).
Typically, we find limits on the lines on the order of $10^{-4}~\mathrm{ph\,cm^{-2}\,s^{-1}}$ at 511\,keV and $10^{-5}~\mathrm{ph\,cm^{-2}\,s^{-1}}$ at 1809\,keV, with V809 Cassiopeia to have the strongest constraints in terms of flux.
Tab.~\ref{tab:preSN_sources} shows the $3\sigma$ upper-limits for the 511\,keV and 1809\,keV flux for all the stars.
As for the most interesting candidate to search for the 1809\,keV line, $\gamma^2$ Velorum \citep{Oberlack1996_26Al, Pleintinger2020_PhD}, any red super giant might be of interest as massive star winds would eject \nuc{Al}{26}.
The mass yield is calculated as follows:
\begin{equation}
    M_{\mathrm{Al}} = 4\pi d^2 \times F_{\mathrm{Al}} \times m_{\mathrm{Al}} \times \tau_{\mathrm{Al}},
\end{equation}
where, $d$ is the distance to the star, $F_{\mathrm{Al}}$ is the expected \nuc{Al}{26} flux, $m_{\mathrm{Al}} = 26~\mathrm{g/mol}$ is the atomic mass of ~\nuc{Al}{26}, and $\tau_{\mathrm{Al}} = 1.04~\mathrm{Myr}$ is the lifetime of ~\nuc{Al}{26}.
The respective ~\nuc{Al}{26} mass yield expected from each star based on the 1809\,keV flux upper limit is also listed in Tab.~\ref{tab:preSN_sources}.
For the closest star in our sample, Spica, with a distance of 77\,pc and a mass of $\sim$12\,M$_\odot$, we find an upper limit on the 1809\,keV line flux of $1.2\times10^{-4}~\mathrm{ph\,cm^{-2}\,s^{-1}}$, which would correspond to an instantaneous \nuc{Al}{26} mass of $6\times10^{-5}~\mathrm{M}_\odot$.
This is more than two orders of magnitude above the expectations from massive star evolution models \citep[e.g.,][]{2012A&A...537A.146E,2018ApJS..237...13L}.
As for the most massive stars in our sample, Rigel ($\sim$21\,M$_\odot$) and Monoceros A ($\sim$29.1\,M$_\odot$), the upper limit on the \nuc{Al}{26} mass is around $1.1\times10^{-4}~\mathrm{M}_\odot$ and $9.0\times10^{-4}~\mathrm{M}_\odot$.
Wind yields of red super giants in the mass range of 20--30\,M$_\odot$ would be found around $5 \times 10^{-7}$--$5 \times 10^{-5}\,\mathrm{M_\odot}$, so that current MeV observations are approaching the interesting region of massive star evolution models.
For the 511\,keV line in the case of Betelgeuse, we found the energy bin from 508--514\,keV is not consistent with zero.
We attribute this to an incomplete modelling of the diffuse emission with the smooth description from \citet{Siegert2016_511}.
In addition, individual sources with such a strong 511\,keV line may be unphysical as their only source would be \nuc{Al}{26} for which we would explain about 41\% of the 1809\,keV line flux at 511\,keV.

\section{Conclusion and outlook}\label{sec:conclusion}
In this work, we present new constraints on ALPs coupled to both electrons and photons, using 22 years of INTEGRAL/SPI observations of 18 nearby red super giants.
	The hot dense stellar interiors provide perfect conditions for the production of ALPs through Compton, Primakoff and to a lesser extent bremsstrahlung processes, which are transformed back to photons in the Galactic magnetic field in the direction towards the star.
    Extending previous studies that focused on the hard X-ray emission from the direction of Betelgeuse, we perform a combined $\gamma$-ray analysis in the $20 - 2000$~keV range across a diverse sample of evolved massive stars.
    We compiled individual limits from a set of 18 red super giants and conducted a Bayesian hierarchical model to constrain the ALP couplings in a joint fit.
    We find no significant ALP-induced emission from any individual source.
    This allows us to set stringent 95\% C.I upper limits on the product $g_{a\gamma} \times g_{ae}$ as a function of ALP mass $m_a$.
    The best case constraints from our full sample analysis improve upon previous limits by up to a factor of $\sim 25$, and reach sensitivity down to $g_{a\gamma} \times g_{ae} = 8 \times 10^{-27}~\mathrm{GeV}^{-1}$ for ultra-light ALPs with $m_a \leqq 10^{-11}$~eV.
    We also explore a conservative scenario accounting for uncertainties in stellar evolutionary stages, assigning one star at a time to a more advanced burning phase while keeping others in earlier phases.
    This yields a robust constraint range of $g_{a\gamma} \times g_{ae} = (0.27 - 1.25) \times 10^{-24}~\mathrm{GeV}^{-1}$.
    Finally, we also look at a Primakoff only emission scenario and constrain the ALP-photon coupling to be in the range $g_{a\gamma} = (0.13 - 1.26)\times10^{-11}~\mathrm{GeV}^{-1}$.
    Our constraints are $\sim 2-3$ orders of magnitude better than the bounds obtained from CAST for Solar ALPs \citep[e.g.,][]{barth2013cast}.
	Furthermore, we also improve on the limits predicted by \textit{Chandra}'s study of conversion in magnetic white dwarfs \citep[e.g.,][]{dessert2022ChandraMWD}. 
    Our results provide one of the most stringent limits to date on ALP couplings in this mass range using $\gamma$-ray data from evolved stars.
    Without modelling the intrinsic astrophysical emission mechanisms in entire galaxies \citep{Ning:2025tit}, estimating ALPs only contribution might be misleading, and could result in overly optimistic limits.
    This was shown in contemporary work about diffuse emission from dark matter in the Milky Way \citet{Berteaud2022_SPI_PBH, siegert2024sub, Siegert2022_MWdiffuse, Calore2023_lightDM}.
    Future improvements may come from more detailed stellar modelling, better understanding of Galactic magnetic field structures, and next-generation $\gamma$-ray instruments with improved sensitivity and resolution, such as NASA's Compton Spectrometer and Imager (COSI) Small Explorer mission which will be launched in 2027~\citep{Tomsick2019_COSI, Tomsick2021_COSI}.
    This work demonstrates the power of multi-source analyses in probing the ALP parameter space and contributes a significant step forward in the search for new physics beyond the Standard Model.

\appendix
\section{Extended energy range}\label{sec:appendix}
\begin{figure}[ht!]
	\includegraphics[width=\linewidth]{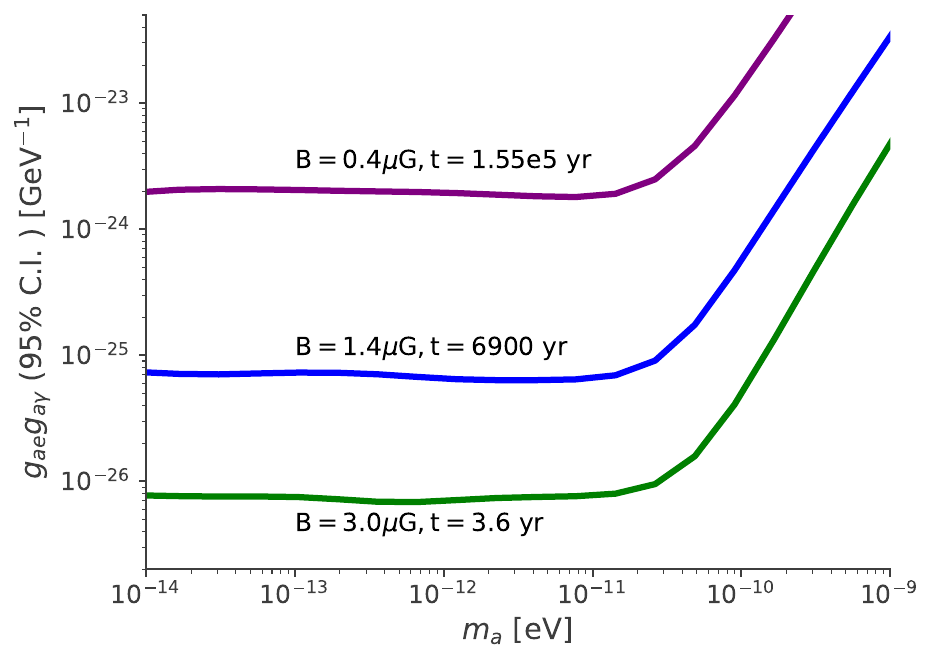}
	\centering
	\caption{The 95\% C.I. upper limits of $g_{ae} \times g_{a\gamma}$ as a function of ALP mass $m_a$ for the combined analysis of all 18 stars for the energy range $20 - 2000~\mathrm{keV}$. The solid blue line shows the upper limit for $\mathrm{t}_{\mathrm{cc}} = 6900$~yr, assuming a representative value of $B_T = 1.4~\mu\mathrm{G}$. The solid purple line shows the upper limit for the most conservative ($B_T = 0.4~\mu\mathrm{G}$ and $t_{cc} = 1.55 \times 10^5~\mathrm{yr}$) and the green line for the most optimistic case ($B_T = 3.0~\mu\mathrm{G}$ and $t_{cc} = 3.6~\mathrm{yr}$). Note that the limits for the extended energy range are identical to the ones shown in Fig.~\ref{fig:all_stars_limits} as the ALP spectrum peaks around 500 keV or below.}
	\label{fig:all_stars_limits_energy_extended}
    \end{figure}

    Fig.~\ref{fig:all_stars_limits_energy_extended} shows the 95\% C.I. for an extended energy range of $20 - 2000$~keV of three different stellar models as quoted.
    The limits obtained for this extended energy range are essentially identical to those obtained for the energy range $20 - 600$~keV.
    This is expected as the ALP spectrum peaks around/below $550$~keV depending on the stellar model, and thus higher energies do not significantly affect the ALP constraints. 

\begin{acknowledgements}
Saurabh Mittal acknowledges support by the  {Bundesministerium für Wirtschaft und Energie via the Deutsches Zentrum für Luft- und Raumfahrt (DLR)} under Contract No. {50\,OO\,2219}. Laura Eisenberger acknowledges support by the Bundesministerium für Wirtschaft und Energie via the Deutsches Zentrum für Luft- und Raumfahrt (DLR) under Contract No. {50\,OR\,2413} and is grateful for the support of the {Stu\-di\-en\-stif\-tung des Deu\-tschen Vol\-kes}. Dimitris Tsatsis acknowledges support from the {DFG/LIS} project {SI\,2502/6-1, project number 551127478}.
This article is based on work from COST Action COSMIC WISPers CA21106, supported by COST (European Cooperation in Science and Technology).
The work of AM and AL  was partially supported by the research grant number 2022E2J4RK "PANTHEON: Perspectives in Astroparticle and Neutrino THEory with Old and New messengers" under the program PRIN 2022 (Mission 4, Component 1, CUP I53D23001110006) funded by the Italian Ministero dell'Universit\`a e della Ricerca (MUR) and by the European Union – Next Generation EU.
This work is (partially) supported by ICSC – Centro Nazionale di Ricerca in High Performance Computing, Big Data and Quantum Computing, funded by European Union–NextGenerationEU.
PC is supported by the Swedish Research Council under contract 2022-04283.
\end{acknowledgements}


\bibliographystyle{aa} 
\bibliography{saurabh} 

\end{document}